# Field measurements demonstrate distinct initiation and cessation thresholds governing aeolian sediment transport flux

Raleigh L. Martin*, Department of Atmospheric and Oceanic Sciences, University of California, Los Angeles, CA 90095

Jasper F. Kok, Department of Atmospheric and Oceanic Sciences, University of California, Los Angeles, CA 90095

**Key Points**
1. We provide the first field-based evidence for separate fluid and impact thresholds in aeolian saltation
2. Saltation occurrence is mediated by both thresholds, but fluid (impact) threshold dominates during infrequent (near-continuous) transport
3. Both thresholds are important for high-frequency saltation prediction, but long-term aeolian fluxes are governed mostly by impact threshold

*raleighmartin@gmail.com

## Abstract
Wind-blown sand and dust models depend sensitively on the threshold wind stress. However, laboratory and numerical experiments suggest the coexistence of distinct "fluid" and "impact" thresholds for the initiation and cessation of aeolian saltation, respectively. Because aeolian transport models typically use only the fluid threshold, existence of a separate lower impact threshold complicates the prediction of wind-driven transport. Here, we derive the first field-based estimates of distinct fluid and impact thresholds from high-frequency saltation measurements at three field sites. Our measurements show that, when saltation is mostly inactive, its instantaneous occurrence is governed primarily by wind exceedance of the fluid threshold. As saltation activity increases, so too does the relative importance of the impact threshold, until it dominates under near-continuous transport conditions. Although both thresholds are thus important for high-frequency saltation prediction, we find that the time-averaged saltation flux is primarily governed by impact threshold.

## 1. Introduction
Determining the threshold wind shear stress for the occurrence of wind-driven ("aeolian") sand transport has been a central challenge for studies of planetary, coastal, and desert aeolian processes [e.g., *Bagnold*, 1941; *Iversen and White*, 1982; *Shao and Lu*, 2000]. In desert and semi-arid environments, the aeolian saltation threshold regulates the frequency of topsoil erosion and mineral dust emission [e.g., *Rice et al.*, 1999]. Where sand dunes and ripples are present, the saltation threshold governs the frequency of migration of these bedforms [e.g., *Fryberger et al.*, 1979]. Recent aeolian transport studies indicate that sand flux scales linearly [e.g., *Ho et al.*, 2011; *Martin and Kok*, 2017] with wind stress in excess of the saltation threshold, so shifts in the presumed threshold value can substantially change predictions of the total sand saltation flux and associated dust emissions [e.g., *Sherman et al.*, 2013; *Kok et al.*, 2014; *Webb et al.*, 2016].

Uncertainty in threshold is therefore a major issue for studies relating aeolian transport observations to atmospheric conditions on Earth [e.g., *Lindhorst and Betzler*, 2016], Mars [e.g., *Bridges et al.*, 2012; *Ayoub et al.*, 2014], and other planetary surfaces [e.g., *Lorenz and Zimbelman*, 2014].

Despite the central importance of the saltation threshold in predicting sand and dust fluxes, there remains a lack of agreement over the best way to model or even measure this threshold [*Barchyn and Hugenholtz*, 2011]. Predictive equations for saltation (and the resulting dust emission) usually include a single threshold value [*Barchyn et al.*, 2014b], traditionally the "fluid" or "static" threshold shear stress $\tau_{ft}$ for initiating saltation transport solely by aerodynamic forces [*Bagnold*, 1941; *Iversen and White*, 1982; *Marticorena and Bergametti*, 1995]. However, theory and measurements suggest the presence of a separate, lower "impact" or "dynamic" threshold shear stress $\tau_{it}$ required to sustain saltation through saltator impacts with the soil bed. This impact threshold has been hypothesized to equal the rate of momentum dissipation at the surface [*Owen*, 1964], which is controversial [*Kok et al.*, 2012], but for which there is now limited experimental support [*Walter et al.*, 2014].

Based on the role of the impact threshold in the steady-state saltation momentum balance, most recent saltation models use impact threshold alone as the zero-intercept value for the saltation flux law [*Ungar and Haff*, 1987; *Creyssels et al.*, 2009; *Martin and Kok*, 2017]. However, recent studies have argued for the need to simulate the path-dependence of saltation flux responses to turbulent wind fluctuations around both impact and fluid thresholds [*Kok*, 2010a, 2010b]. The ratio of impact and fluid thresholds governing this hysteresis depends primarily on the particle-fluid density ratio, which determines the relative contributions of particle impacts and direct fluid lifting to particle entrainment [*Kok*, 2010b; *Pähtz and Durán*, 2017]. On Earth, the experimentally [*Bagnold*, 1937; *Chepil*, 1945; *Iversen and Rasmussen*, 1994] and numerically [*Kok*, 2010b] predicted ratio of impact and fluid threshold shear velocities $u_{*it}/u_{*ft}$ is approximately 0.82, whereas $u_{*it}/u_{*ft}$ is predicted to be as low as 0.1 on Mars [*Kok*, 2010a].

Though numerical and experimental studies predict fluid and impact thresholds, field studies have not yet confirmed the existence of these two separate thresholds, let alone determined which should be used for modeling sand and dust flux. One possible approach to field-based determination of fluid and impact thresholds is to directly examine the correlation between the wind speed and the occurrence of saltation onset and cessation events. However, the signal of dual thresholds revealed by such an approach is typically overwhelmed by the noise in highly variable wind and saltation measurements [*Barchyn and Hugenholtz*, 2011; *Martin et al.*, 2013]. Comparisons of saltation flux measurements to time series of wind speed [e.g., *Wiggs et al.*, 2004b; *Davidson-Arnott et al.*, 2005; *Davidson-Arnott and Bauer*, 2009] or momentum flux [e.g., *Sterk et al.*, 1998] display a poor correlation at short time scales, due to spatial separation between saltation and wind measurements [*Baas*, 2008], variability in surface grain configurations [*Nickling*, 1988; *Li et al.*, 2008], small-scale fluctuations in the turbulent winds driving saltation [*Carneiro et al.*, 2015], and decoupling of saltation and wind systems [*Paterna et al.*, 2016]. Therefore, such a deterministic correlation-based approach to measuring fluid and impact thresholds in the field is difficult.

Here, we instead adopt a statistical approach to determine distinctive fluid and impact thresholds in aeolian saltation from extensive high-frequency field measurements of aeolian transport at three field sites. In the following sections, we develop a "dual threshold hypothesis" for how fluid and impact thresholds determine the effective threshold governing the frequency of saltation activity (Section 2), describe methods for calculating these saltation activities and effective thresholds from field measurements (Section 3), use these calculations to test the dual threshold hypothesis and derive field-based estimates of separate fluid and impact thresholds (Section 4), discuss the role of fluid and impact thresholds in modeling sand and dust fluxes (Section 5), and conclude with a summary of this work (Section 6).

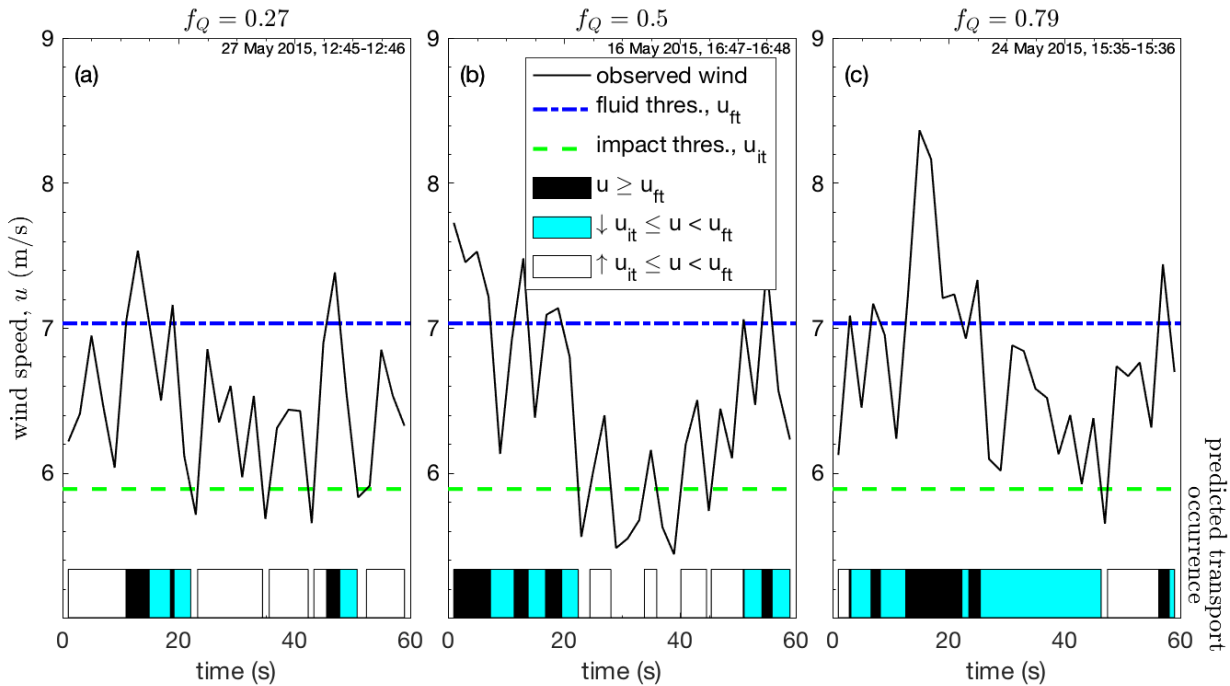

**Figure 1.** Illustration of variations in wind speed $u$ and predicted saltation occurrence for time intervals with (a) low ($f_Q = 0.27$), (b) medium ($f_Q = 0.5$), and (c) high ($f_Q = 0.79$) observed transport activity at the Oceano field site (introduced in Section 3). Blue and green dashed lines indicate respective fluid and impact threshold wind speeds, $u_{ft}$ and $u_{it}$. Black bars at the bottom of each panel refer to times when $u \geq u_{ft}$. Both cyan and white bars refer to "intermittent zone" times when $u_{it} \leq u < u_{ft}$, but they are distinguished between times approached from above $u_{ft}$ (cyan) versus from below $u_{it}$ (white). Black and cyan bars thus indicate times of predicted transport occurrence. As saltation activity $f_Q$ increases from panel a to c, the intermittent zone is increasingly approached from above and decreasingly from below.

## 2. Theory

We expect the occurrence of aeolian saltation transport to be governed by both the fluid and impact thresholds, based on the relative importance of transport initiation versus cessation. In this section, we describe how, over time intervals of intermittent saltation that include many threshold crossings, the combined contributions of $\tau_{ft}$ and $\tau_{it}$ should produce an intermediate "effective" threshold stress $\tau_{th}$ that varies systematically with the fraction of time $f_Q$ that saltation is active. To demonstrate this theory, we utilize sample time series of wind speed $u$ and saltation activity $f_Q$ from our field campaigns. We will describe methods for obtaining these time series in Section 3.

## 2.1. Regulation of saltation occurrence by fluid and impact thresholds
To illustrate the role of dual thresholds in regulating the occurrence of saltation, we consider three sample time series (Fig. 1) of near-surface (anemometer height $z_U \approx 0.5$ m) horizontal wind speed $u(t)$ straddling fluid and impact threshold wind speeds, $u_{ft}$ and $u_{it}$ respectively, corresponding to $\tau_{ft}$ and $\tau_{it}$. (Methods for obtaining $u(t)$, $u_{ft}$, and $u_{it}$ will be described in Section 3.) When $u \geq u_{ft}$, we unambiguously expect saltation transport to occur; conversely, when $u < u_{it}$, transport should not occur. Ambiguity in prediction of saltation occurrence arises in cases where $u_{it} \leq u < u_{ft}$. In this "intermittent zone," saltation occurrence should depend on whether saltation transport was initiated (i.e., $u \geq u_{ft}$) more recently than it was terminated (i.e., $u < u_{it}$) [*Kok*, 2010b].

Over time intervals covering multiple threshold crossings, we expect the frequency of saltation activity $f_Q$ to partially depend on the fraction of intermittent zone winds (i.e., $u_{it} \leq u < u_{ft}$) that are approached from below ($u < u_{it}$) versus from above ($u \geq u_{ft}$). Fig. 1 illustrates the nature of these intermittent zone events for three distinctive cases of measured infrequent (panel a: $f_Q = 0.27$), moderate (panel b: $f_Q = 0.5$), and frequent (panel c: $f_Q = 0.79$) saltation activity. (Methods for calculating $f_Q$ from measurements will be described in Section 3.) When the intermittent zone is mostly approached from below (i.e., from a state of non-transport), saltation is mostly limited by the occurrence of initiation events, for which $u \geq u_{ft}$ (Fig. 1a). As such, near the no saltation limit $f_Q \to 0$, saltation activity is controlled primarily by wind exceedance of the fluid threshold. In the contrasting case when the intermittent zone is mostly approached from above ($u \geq u_{ft}$) (i.e., from a state of transport), saltation will mostly be sustained as long as $u \geq u_{it}$ (Fig. 1c). Consequently, near the continuous saltation limit $f_Q \to 1$, saltation occurrence is controlled primarily by wind exceedance of the impact threshold. Otherwise, for cases of intermittent zone winds originating equally from starting points above $u_{ft}$ and below $u_{it}$ (Fig. 1b), expectations for saltation occurrence will be somewhere between these two end-member cases. In general, the examples in Fig. 1 suggest that, with an increasing fraction of intermittent zone winds originating from $u \geq u_{ft}$ relative to $u < u_{it}$, saltation occurrence will move away from the $f_Q \to 0$ limit of $u_{ft}$ control toward the $f_Q \to 1$ limit of $u_{it}$ control.

## 2.2. Dual threshold hypothesis for saltation activity
The three cases presented in Fig. 1 suggest that increasing $f_Q$ corresponds to saltation occurrence being increasingly controlled by the impact rather than the fluid threshold. Based on this observation, we consider how a statistically-defined "effective" threshold wind shear stress $\tau_{th}$, which refers to the average threshold wind stress above which saltation is expected to occur [e.g., *Stout and Zobeck*, 1997], reflects the relative contributions of fluid and impact threshold stresses, $\tau_{ft}$ and $\tau_{it}$, with increasing $f_Q$. In particular, we propose a "dual threshold hypothesis", in which $\tau_{th} = \tau_{ft}$ in the initiation-limited rare transport case ($f_Q \to 0$), $\tau_{th} = \tau_{it}$ in the cessation-limited continuous transport case ($f_Q \to 1$), and $\tau_{th}$ decreases linearly with $f_Q$ between these two limits:
$$\tau_{th} = f_Q \tau_{it} + (1 - f_Q) \tau_{ft}. \tag{1}$$
This hypothesis is consistent with *Schönfeldt* [2004], who predicted an analogous gradual decrease in effective threshold with increasing mean wind speed (rather than saltation activity) in stochastic simulations of wind speed fluctuating around fluid and impact thresholds. If correct, Eq. 1 offers a way to determine the impact and fluid thresholds from measurements of effective threshold and saltation activity. Conversely, Eq. 1 could also allow for prediction of saltation activity from wind time series and known impact and fluid thresholds.

**Table 1.** Threshold values for primary analysis with $\delta t$ = 2s averaging interval and $\Delta t$ = 1 minute analysis interval. Median, 10$^{th}$ percentile, and 90$^{th}$ percentile grain diameters of surface particles at each site ($d_{50}$, $d_{10}$, and $d_{90}$) are also included for reference.

| Site | Median grain diam., $d_{50}$ (mm) | 10$^{th}$ pctile. grain diam. $d_{10}$ (mm) | 90$^{th}$ pctile. grain diam., $d_{90}$ (mm) | Fluid threshold stress, $\tau_{ft}$ (Pa) | Impact threshold stress, $\tau_{it}$ (Pa) | Threshold ratio, $u_{*,it}/u_{*,ft}$ | Threshold from flux-law fit, $\tau_{th,flux}$ (Pa) |
|---|---|---|---|---|---|---|---|
| Jericoacoara | 0.526 ±0.037 | 0.097 ±0.012 | 0.847 ±0.037 | 0.168 ±0.004 | 0.111 ±0.002 | 0.813 ±0.018 | 0.135 ±0.015 |
| Rancho Guadalupe | 0.533 ±0.026 | 0.219 ±0.035 | 0.839 ±0.034 | 0.147 ±0.006 | 0.110 ±0.002 | 0.863 ±0.027 | 0.110 ±0.021 |
| Oceano | 0.398 ±0.070 | 0.190 ±0.032 | 0.650 ±0.075 | 0.125 ±0.001 | 0.088 ±0.001 | 0.837 ±0.007 | 0.094 ±0.006 |

## 3. Methods
### 3.1. Field deployments
To evaluate our dual threshold hypothesis that effective threshold is partitioned between fluid and impact thresholds depending on the saltation activity (Eq. 1), we analyze here simultaneous high-frequency measurements of active saltation and wind at three field sites: Jericoacoara, Ceará, Brazil (2.7969°S, 40.4837°W); Rancho Guadalupe, California, U.S.A. (34.9592°N, 120.6431°W); and Oceano, California, U.S.A. (35.0287°N, 120.6277°W). All field sites contain mostly flat, unvegetated, sand-covered surfaces, with distinctive sediment size distributions. Median, 10$^{th}$, and 90$^{th}$ percentile grain diameters ($d_{50}$, $d_{10}$, and $d_{90}$, respectively), determined through Camsizer optical grain-size analysis [*Jerolmack et al.*, 2011] of multiple surface samples collected at each field site, are listed in Table 1.

During each field campaign, multiple (3-9) Wenglor optical sensors [*Barchyn et al.*, 2014a] at heights from the bed surface up to ≈0.3 m counted saltating particles (at 25 Hz), which we convert to vertically-integrated saltation particle count rates $N$ (Fig. 2a). A sonic anemometer at height $z_U \approx 0.5$ m [*Martin et al.*, 2017] measured wind velocity $u$ (25 Hz at Jericoacoara and Rancho Guadalupe, 50 Hz at Oceano) (Fig. 2d). Though our field deployments (described further in *Martin et al.* [2017]) included anemometers at multiple heights, we choose here to use measurements from only the lowest anemometer (i.e., $z_U \approx 0.5$ m) at each site, because we expect these measurements to be most representative of wind fluctuations at the sand surface. Wind measurements described in this analysis are thus vertically offset from saltation measurements by ≈0.2-0.5 m. At Jericoacoara and Rancho Guadalupe, wind measurements are additionally separated from saltation measurements by $\approx 1$ m in the spanwise direction, whereas they are spanwise co-located at Oceano [*Martin et al.*, 2017].

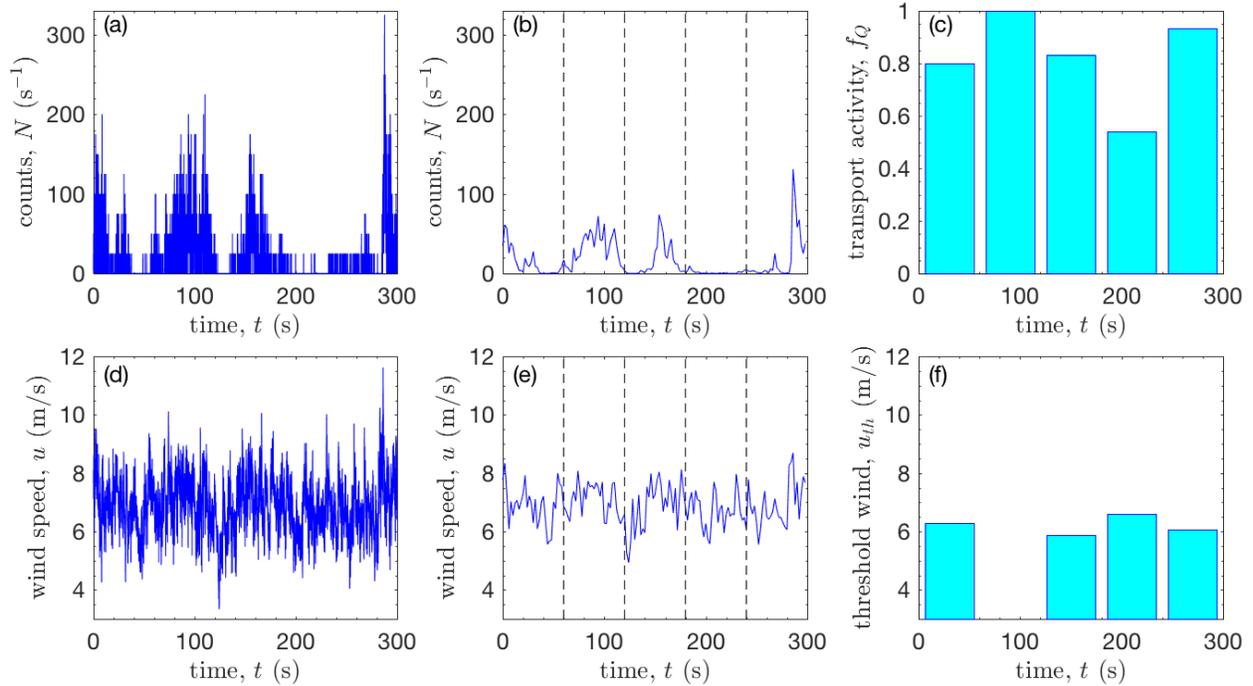

**Figure 2.** Sample measurements (Rancho Guadalupe, 24 March 2015, 13:31-13:36) to illustrate methods for calculating saltation transport activity $f_Q$ and effective threshold wind speed $u_{th}$. (a) Time series of total particle counts rate $N$. (b) $\delta t = 2$s interval-averaged $N$ time series. Dashed lines indicate $\Delta t = 1$ minute analysis intervals. (c) Transport activities $f_Q$, calculated as fraction of $\delta t$ increments in each $\Delta t$ for which $N > 0$, and corrected for false negatives. (d) Time series of streamwise wind speed $u$. (e) $\delta t = 2$s interval-averaged $u$. Dashed lines again indicate $\Delta t$. (f) Resulting values of threshold wind speed $u_{th}$ calculated by Eq. 2 from corresponding $f_Q$ and wind speed distributions $\Phi_u$ for each $\Delta t$. $u_{th}$ is undefined for $t = 60$-$120$s when $f_Q = 1$.

### 3.2. Calculating effective thresholds

To calculate saltation activity $f_Q$ and effective threshold wind speed $u_{th}$, we apply $\delta t = 2$s interval averaging to saltation (Fig. 2b) and wind (Fig. 2e) time series, then we subdivide these data into $\Delta t = 1$ minute analysis intervals. Within each analysis interval, we calculate saltation activity $f_Q$ as the fraction of $\delta t$ increments within $\Delta t$ for which $N$ is nonzero (Fig 2c). Due to counting uncertainty for the small sampling volume of Wenglor sensors counting the passage of individual particles over short time intervals, we apply a slight correction to $f_Q$ to account for the possibility of false negatives (see Appendix A). For each $\Delta t$, we then calculate effective threshold wind speed $u_{th}$ by applying the "Time Frequency Equivalence Method" [*Stout and Zobeck*, 1997; *Stout*, 2004; *Wiggs et al.*, 2004b]:

$$u_{th} = \Phi_u(1 - f_Q), \qquad (2)$$

where $\Phi_u(1 - f_Q)$ is the value in the cumulative distribution of wind speeds $\Phi_u$ corresponding to the time fraction of inactive saltation, $1 - f_Q$ (Fig. 2f). When calculating $\Phi_u$ for each $\Delta t$, we use the $\delta t$-averaged $u$ values, in correspondence with our methods for calculating $f_Q$. We choose the averaging interval $\delta t = 2$s based on the typical response time of saltation to turbulent wind fluctuations [e.g., *Anderson and Haff*, 1988; *McEwan and Willetts*, 1991; *Ma and Zheng*, 2011], and we choose the analysis interval $\Delta t = 1$ minute to represent the typical oscillation period for large-scale structures in an atmospheric boundary layer (ABL) [e.g., *Guala et al.*, 2011]. We note here that, due to the spatial separation of wind and saltation measurements, we cannot directly relate individual high-frequency wind gusts to individual occurrences of saltation. Instead, we assume, based on the correspondence of low-frequency wind fluctuations across a wide range of heights [e.g., *Marusic et al.*, 2010], that measured wind fluctuations over $\Delta t$ are statistically representative of the wind experienced by the measured saltating particles.

To account for statistical variability and facilitate uncertainty estimation, we group effective threshold wind speed $u_{th}$ values computed over individual $\Delta t$ intervals into $f_Q$ bins (see Appendix B). We then convert binned $u_{th}$ values into effective threshold shear velocities $u_{*,th}$ and stresses $\tau_{th}$ by the law-of-the-wall and the standard $\tau$-$u_*$ relationship:

$$u_{th} = \frac{u_{*,th}}{\kappa} \ln\left(\frac{z_U}{z_0}\right), \qquad (3)$$

$$\tau_{th} = \rho_f u_{*,th}^2, \qquad (4)$$

where $z_0$ is aerodynamic roughness height, $\kappa \approx 0.4$ is the von Karman parameter, and $\rho_f$ is air density determined from the mean temperature at each site [*Martin and Kok*, 2017]. We apply standard error propagation techniques to estimate uncertainties in resulting calculated $u_{*,th}$ and $\tau_{th}$ values (Appendix B).

### 3.3. Applicability of the law-of-the-wall

There are three possible issues in applying the law-of-the-wall (Eq. 3) to convert from threshold wind speeds to shear velocities and shear stresses. First, the law-of-the-wall is only strictly valid for unidirectional winds within a neutrally stable ABL [e.g., *Frank and Kocurek*, 1994], but we frequently observe unstable wind conditions. Second, the law-of-the-wall requires time averaging of the wind speed over a sufficiently long time period to capture the full range of turbulence variability [e.g., *van Boxel et al.*, 2004], but we apply Eq. 3 to threshold wind speeds $u_{th}$ obtained from quasi-instantaneous values within the wind speed distribution, which do not represent time averages. Third, measured roughness height is known to change systematically with saltation intensity [e.g., *Sherman*, 1992], but we use a constant $z_0$ in Eq. 3. We address each of these issues in the subsections below.

#### *3.3.1. Law-of-the-wall: stability and wind direction*

To address the need for unidirectional and neutrally stable conditions for application of the law-of-the-wall, we note that, during all periods of active saltation for which we perform threshold calculations, measured wind directions (i.e., $\theta$) deviate by less than 20° from the mean sediment-transporting wind. Furthermore, stability parameter values (i.e., $|z/L|$) are always less than 0.2 during active saltation (See *Martin and Kok* [2017] for further explanation of these calculations). For such unidirectional and neutrally stable winds during active saltation, law-of-the-wall and Reynolds stress-based methods for computing the shear velocity are roughly equivalent [*Salesky et al.*, 2012]. This contrasts with observed wide variation in $\theta$ and $|z/L|$ during non-saltation conditions [*Martin and Kok*, 2017], but our threshold calculations necessarily exclude such non-saltation time intervals.

### *3.3.2. Law-of-the-wall: averaging time*
To address the issue of time averaging, we note that independent measurements [*Namikas et al.*, 2003] have demonstrated the convergence of law-of-the-wall profiles over time periods as small as 10 seconds, shorter than our chosen $\Delta t = 1$ minute. Furthermore, we make our threshold calculations only for anemometers mounted close to the bed surface (i.e., $z_U \approx 0.5$m) that experience moderate to strong mean winds associated with saltation (i.e., $u \gtrsim 5$ m/s), suggesting reasonably rapid convergence of wind statistics [*van Boxel et al.*, 2004].

However, we apply the law-of-the-wall not to the mean wind over the $\Delta t = 1$ minute analysis interval but to a single value ($u_{th}$) within a distribution $\Phi_u$ of winds over $\Delta t$; thus, the required measurement period for application of the law-of-the-wall may be longer than our chosen $\Delta t$. To test for this possibility, we perform sensitivity analyses to assess possible changes in $u_{*,th}$ and $\tau_{th}$ with varying $\Delta t$. We find no systematic effect of varying $\Delta t$, suggesting that our choice of $\Delta t = 1$ minute is sufficient for application of the law-of-the-wall. We address this issue further in the Discussion.

### *3.3.3. Law-of-the-wall: roughness height*
Now, we address the selection of and possible saltation-induced variation in roughness height $z_0$ in Eq. 3. Because roughness height is known to increase with saltation intensity [e.g., *Sherman*, 1992], it is generally preferable to calculate $z_0$ from wind profiles measured during non-saltation conditions. However, due to the occurrence of unstable wind profiles during non-saltation conditions causing deviations from logarithmic profiles (Sec. 3.3.1), we find that such measurements of $z_0$ display large variability over many orders of magnitude.

To avoid the problem of unstable profiles, we instead consider roughness heights measured during stronger winds, when near-surface wind profiles are closer to neutral stability. However, such stronger winds are associated with the occurrence of saltation and with a general increase in roughness height [*Sherman*, 1992]. We therefore refer to these saltation-influenced roughness values as "effective" roughness heights $z_s$ to distinguish them from the purely aerodynamic roughness $z_0$. As with $z_0$, the value for $z_s$ can be determined by manipulating the law-of-the-wall:

$$z_s = z_U \exp\left(-\frac{\kappa u}{u_{*,Re}}\right), \tag{5}$$

where $u$ here refers to the mean streamwise wind speed for the anemometer mounted at height $z_U$, and $u_{*,Re}$ is the shear velocity determined by the Reynolds stress method over a 30 minute interval [*Martin and Kok*, 2017]. We choose to use 30-minute intervals to calculate $z_s$ for comparison to $u_{*,Re}$ (instead of the $\Delta t = 1$ minute analysis intervals applied elsewhere in this paper), because $u_{*,Re}$ is ill-defined at a time scale of 1 minute [*van Boxel et al.*, 2004].

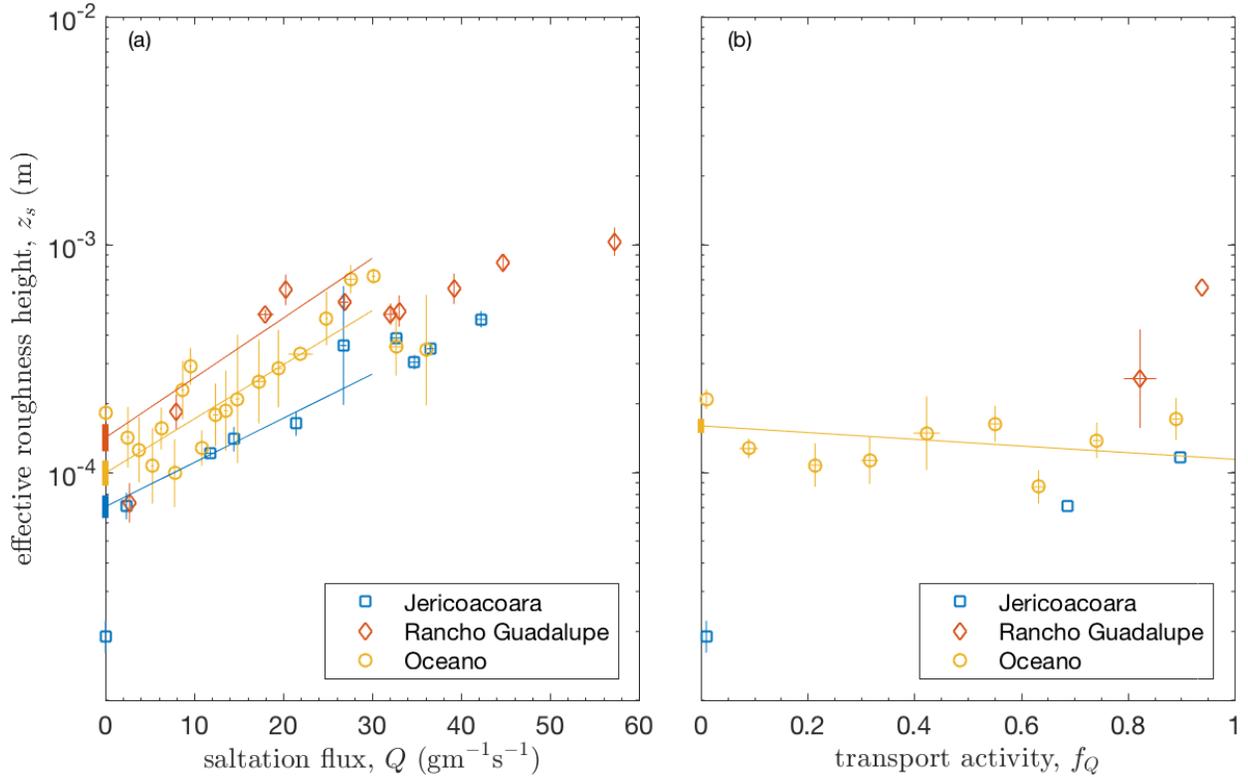

**Figure 3.** (a) Comparison of effective roughness height $z_s$ versus saltation flux $Q$ at each field site. We estimate $z_0$ as the zero-intercept of the linear fit of $\ln(z_s)$ versus $Q$ at each site for $Q \leq 30$ gm$^{-1}$s$^{-1}$, indicated by the solid lines. Resulting calculated $z_0$ values for each site are shown as thick vertical bars at $Q = 0$ gm$^{-1}$s$^{-1}$. The vertical range of these bars corresponds to uncertainties in $z_0$ as determined by the linear fits. (b) Comparison of effective roughness height $z_s$ versus saltation transport activity $f_Q$ for intermittent transport ($0.05 < f_Q < 0.95$). Values are calculated over 30-minute intervals and combined into bins by $f_Q$. Ranges of 30-minute $f_Q$ values at Jericoacoara and Rancho Guadalupe are insufficient for binning and fitting.

Comparing $z_s$ values to saltation fluxes $Q$ measured over corresponding intervals (see *Martin and Kok* [2017] for methods of estimating $Q$), we observe an increasing trend (Fig. 3a) consistent with expectations for saltation-induced roughness [*Sherman*, 1992]. At each site, the logarithm of $z_s$ increases linearly with $Q$ up to $Q \approx 30$ gm$^{-1}$s$^{-1}$. Performing a linear fit to $\ln(z_s)$ versus $Q$ over this range, we estimate $z_0$ as the zero-intercept (i.e., $Q = 0$) value of this fit, yielding values of $z_0 = 0.707 \times 10^{-4}$ m, $1.420 \times 10^{-4}$ m, and $0.993 \times 10^{-4}$ m, for Jericoacoara, Rancho Guadalupe, and Oceano, respectively. Associated uncertainties in log-space, i.e. $\sigma_{\ln(z_0)}$, are 0.115, 0.137, and 0.128, for Jericoacoara, Rancho Guadalupe, and Oceano, respectively. Thus, Fig. 3a demonstrates a way to estimate $z_0$ from the variation in $z_s$ with $Q$. However, it does not answer the question of which roughness height(s) to use for converting from effective threshold wind speeds $u_{th}$ to effective threshold shear velocities $u_{*,th}$ in Eq. 3.

To select the proper roughness heights for Eq. 3, we consider the variation in effective roughness height $z_s$ with saltation transport activity $f_Q$ measured over corresponding 30-minute intervals. This comparison is only possible at Oceano, where 30-minute values of $f_Q$ cover the full possible range of $f_Q$ from 0 to 1. Notably, there appears to be a negligible variation in $z_s$ with $f_Q$ (Fig. 3b). We confirm this lack of trend by performing a linear fit to $\ln(z_s)$ versus $f_Q$. The best fit for this slope, $-0.34 \pm 0.17$, indicates a weak but not statistically significant negative trend. Though such a result seems at odds with the obvious increase in $z_s$ with $Q$ (Fig. 3a), we note that most intermittent transport, i.e., $0.05 < f_Q < 0.95$, corresponds to relatively small saltation fluxes, for which the deviation of $z_s$ away from $z_0$ is negligible. We therefore use our empirically-derived $z_0$ values (Fig. 3a) for all threshold wind speed to shear velocity calculations (Eq. 3).

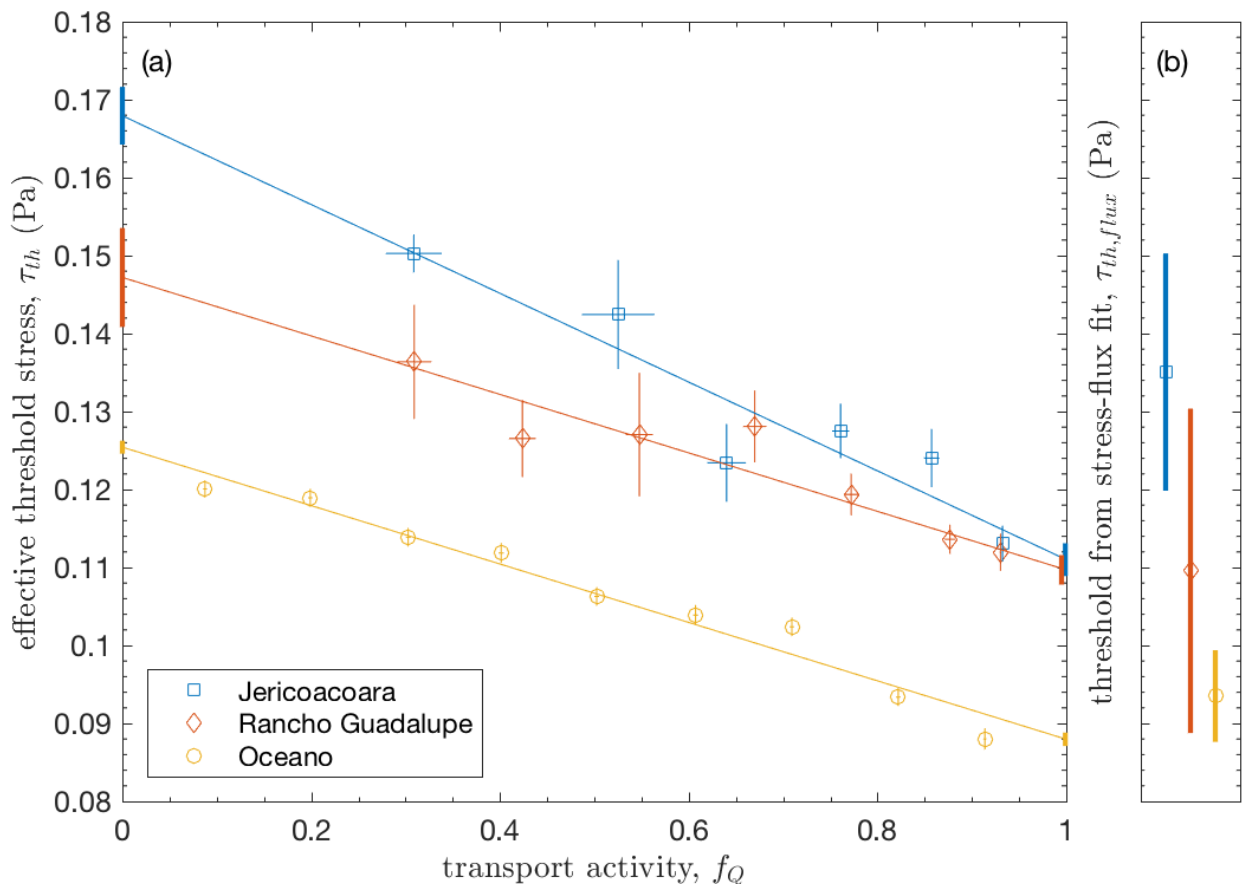

**Figure 4.** (a) Effective threshold stress $\tau_{th}$ versus saltation transport activity $f_Q$. Color-coded lines show least-squares fits (Eq. 6) at each field site. Vertical bars at $f_Q = 0$ and $f_Q = 1$ denote respective estimates of the fluid threshold $\tau_{ft}$ (Eq. 7) and impact threshold $\tau_{it}$ (Eq. 8) from the linear fit. (b) We estimate flux-based thresholds $\tau_{th,flux}$ for each field site by applying Eq. 10 to saltation flux measurements reported in *Martin and Kok* [2017]. Specific values for $\tau_{ft}$, $\tau_{it}$, and $\tau_{th,flux}$ are listed in Table 1. Error bars correspond to 1 standard error.

## 4. Results

To evaluate our hypothesis for dual threshold control of saltation activity and the effective saltation threshold (Eq. 1), we use the methods described in Section 3 to calculate effective threshold stresses and saltation activities at our three field sites. Due to the distinctive soil properties at each field site, we analyze the effective thresholds and saltation activities for each site separately.

### 4.1. Fluid and impact thresholds

As predicted by Eq. 1, we find that the effective threshold stress $\tau_{th}$ decreases linearly with saltation activity $f_Q$ at each of the field sites (Fig. 4a). To quantify this trend, we perform a linear fit to $\tau_{th}$ versus $f_Q$, i.e.,

$$\tau_{th} = a + bf_Q, \tag{6}$$

where $a$ is the fitting intercept and $b$ is the fitting slope. Based on uncertainties in these linear fitting parameters, we find that the observed decline in $\tau_{th}$ with $f_Q$ is extremely unlikely by random chance alone ($p < 10^{-6}$). Based on this linear fit, we also calculate fluid and impact threshold stresses, $\tau_{ft}$ and $\tau_{it}$, from the limiting effective threshold values for no transport ($f_Q \rightarrow 0$) and continuous transport ($f_Q \rightarrow 1$) in Eq. 1, i.e.:

$$\tau_{ft} = a, \tag{7}$$
$$\tau_{it} = a + b. \tag{8}$$

We follow standard error propagation methods [*Bevington and Robinson*, 2003] to determine uncertainties in $\tau_{ft}$ and $\tau_{it}$ from uncertainties in $a$ and $b$ for the linear fit (Appendix B). Calculated threshold values and their uncertainties, illustrated in Fig. 4a and listed in Table 1, are comparable to wind tunnel measured values of $\tau_{ft}$ [*Bagnold*, 1937; *Chepil*, 1945; *Zingg*, 1953; *Fletcher*, 1976; *Kok et al.*, 2012: Fig. 5] and $\tau_{it}$ [*Bagnold*, 1937; *Chepil*, 1945; *Iversen and Rasmussen*, 1994; *Li and McKenna Neuman*, 2012; *Kok et al.*, 2012: Fig. 21] for sediment bed grain sizes similar to those measured at our field sites.

### 4.2. Threshold ratios

Converting fluid and impact threshold stresses to threshold shear velocities by Eq. 4, we calculate threshold ratios as:

$$\frac{u_{*it}}{u_{*ft}} = \sqrt{\frac{\tau_{it}}{\tau_{ft}}}. \tag{9}$$

Based on Eq. 9, we calculate $u_{*,it}/u_{*,ft} = 0.813 \pm 0.018$, $0.863 \pm 0.027$, and $0.837 \pm 0.007$ at Jericoacoara, Rancho Guadalupe, and Oceano, respectively (Table 1). These values are consistent with laboratory measurements [*Bagnold*, 1937; *Chepil*, 1945; *Iversen and Rasmussen*, 1994] and numerical predictions [*Kok*, 2010a] of $u_{*,it}/u_{*,ft} \approx 0.82$ (i.e., $\tau_{it}/\tau_{ft} \approx 0.67$). As with threshold stresses, we compute uncertainties in $u_{*,it}/u_{*,ft}$ by following standard error propagation methods (Appendix B).

## 5. Discussion

Our results provide the first field-based evidence for the existence of separate fluid and impact thresholds in aeolian saltation. Though fluid and impact thresholds for saltation initiation and cessation have long been theorized [e.g., *Bagnold*, 1937; *Kok*, 2010b] and measured in wind tunnel experiments [e.g., *Chepil*, 1945; *Iversen and Rasmussen*, 1994], the difficulty of directly measuring threshold crossings in the field [e.g., *Barchyn and Hugenholtz*, 2011] has limited the ability of past studies to resolve both thresholds. To overcome these limitations, we hypothesized, based on observations (Fig. 1), that the relative contributions of fluid and impact thresholds in controlling saltation occurrence vary systematically with the frequency of saltation transport (Eq. 1). As expected from this dual threshold hypothesis, we observed that the statistically-defined effective threshold stress [*Stout and Zobeck*, 1997; *Stout*, 2004] (Fig. 2) decreases linearly with saltation activity between two end-member cases of fluid threshold dominance in the rare transport limit and impact threshold dominance in the continuous transport limit (Fig. 4a). Based on these limiting effective threshold values, we then calculated distinct fluid and impact threshold stresses at the three field sites, which we found to be consistent with past theory and experiments.

Our observations offer two primary pieces of evidence supporting the combined role of distinctive fluid and impact thresholds in controlling saltation occurrence. First, a systematic decrease in effective threshold with increasing saltation activity (Fig. 4a) is consistent with our dual threshold hypothesis (Eq. 1), which predicts a gradual shift from fluid threshold to impact threshold control with increasing transport activity. Second, our estimated values for impact and fluid threshold (Table 1) are consistent with independent estimates of impact threshold versus grain size and predicted ratios of these values. These pieces of evidence thus also lend support to our statistically-based approach for determining fluid and impact thresholds from field measurements of wind speed and saltation activity.

### 5.1. Potential limitations and sensitivity analyses

Despite this evidence supporting the existence of dual thresholds, we consider aspects of our methodology that could produce an artificial variation in effective threshold with saltation activity. First, our measured thresholds could depend on the selection of averaging intervals $\delta t$ (Fig. 2) [*Stout*, 1998; *Schönfeldt*, 2003; *Wiggs et al.*, 2004b; *Barchyn and Hugenholtz*, 2011]. Saltation is more likely to occur within longer $\delta t$ increments (Fig. 2b), thus increasing $f_Q$ (Fig. 2c). Longer $\delta t$ also decreases the amplitude of wind fluctuations (Fig. 2e), which reduces the range of measured effective thresholds $u_{th}$ (Fig. 2f). To evaluate these $\delta t$ effects, we perform a sensitivity analysis to reinvestigate the $\tau_{th}$ versus $f_Q$ relationship for five values of $\delta t$ ranging over 1-4 seconds, corresponding to typical saltation response times [e.g., *Anderson and Haff*, 1988; *McEwan and Willetts*, 1991; *Ma and Zheng*, 2011]. Fig. 5a shows that, though the $\tau_{th}$ versus $f_Q$ relationship is somewhat affected by variation in $\delta t$, calculated values of $\tau_{it}, \tau_{ft},$ and $u_{*it}/u_{*ft}$ (Table 2) remain broadly consistent with independent measurements [e.g., *Bagnold*, 1937; *Kok*, 2010b] regardless of $\delta t$.

Second, our findings could also depend on the selection of analysis interval $\Delta t$, which partially determines the distribution of wind speeds $\Phi_u$ from which $u_{th}$ values are calculated. In particular, we consider the possibility that our calculated $u_{ft}$ and $u_{it}$ values are merely statistical artifacts of our threshold determination method, arising simply from the respective lower and upper limits of the $\Phi_u$ distribution. If this were the case, increasing $\Delta t$, which typically produces a wider wind speed distribution $\Phi_u$, would widen the separation between calculated $\tau_{ft}$ and $\tau_{it}$ values. However, sensitivity analyses for $\Delta t = 0.5$-30 minutes reveal no systematic trends (Fig.

5b). This insensitivity to $\Delta t$ leads us to reject the possibility that $\tau_{ft}$ and $\tau_{it}$ are merely statistical artifacts of our methodology. Such insensitivity to $\Delta t$ also suggests that our adoption of the law-of-the-wall for conversion from threshold wind speed to threshold shear velocity (Eq. 3) is unaffected by measurement time interval (Sec. 3.3).

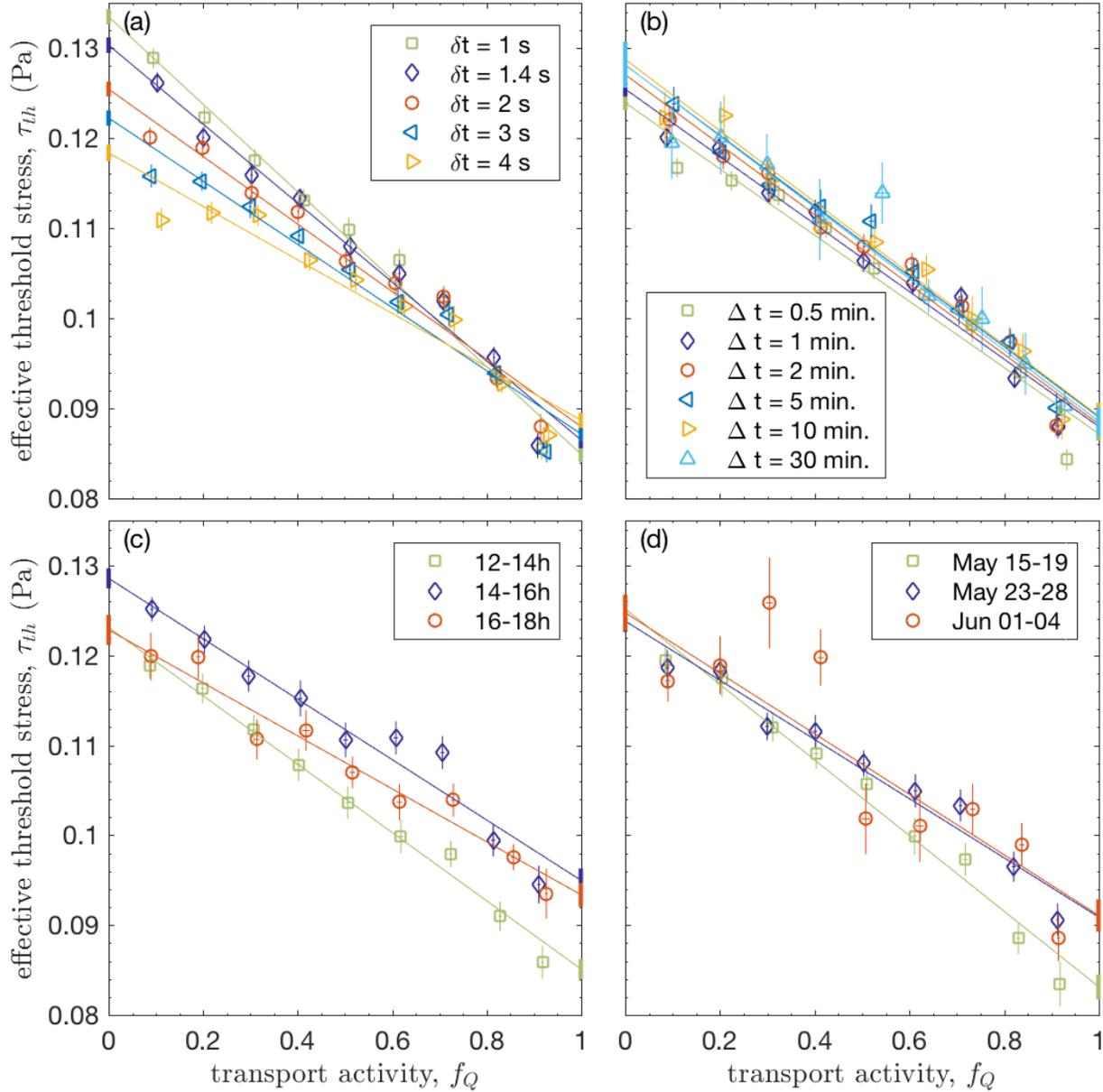

**Figure 5.** Sensitivity analyses comparing effective threshold $\tau_{th}$ and saltation transport activity $f_Q$. We limit our analyses here to the Oceano field site, where the most data are available to explore a wide range of averaging intervals $\delta t$, analysis intervals $\Delta t$, and conditional analyses by time of day and date. For all plots, limiting fluid and impact threshold values and their uncertainties, indicated by vertical bars at $f_Q = 0$ and $f_Q = 1$, respectively, are given in Table 2. (a) $\tau_{th}$ versus $f_Q$ comparison for values of $\delta t$ =1-4 s. Analysis interval $\Delta t$ is fixed at 1 minute

for this analysis. (b) Comparison for six different analysis intervals $\Delta t$ =0.5-30 minutes. Averaging interval $\delta t$ is fixed at 2 s for this analysis. (c) Comparison for three different time periods for the diurnal cycle at Oceano. Here, both averaging interval ($\delta t = 2$s) and analysis interval ($\Delta t = 1$ minute) are fixed. (d) Comparison for three date range segments of the deployment at Oceano, each of which is characterized by a distinctive surface grain size distribution (Table 2). Again, $\delta t$ and $\Delta t$ are fixed.

**Table 2.** Best fit values for sensitivity analyses at Oceano field site for averaging interval $\delta t$ (Fig. 5a), analysis interval $\Delta t$ (Fig. 5b), diurnal time interval (Fig. 5c), and date range (Fig. 5d). Uncertainties correspond to linear fitting uncertainty, which accounts for the uncertainty in the individual data points used for this fitting.

|  | Fluid threshold, $\tau_{ft}$ (Pa) | Impact threshold, $\tau_{it}$ (Pa) | Threshold ratio, $u_{*,it}/u_{*,ft}$ |
|---|---|---|---|
| **Averaging interval $\delta t$ (s)** | *For sensitivity analysis (Fig. 5a), $\Delta t$ held constant at 1 minute* | | |
| 1 | 0.133±0.001 | 0.085±0.001 | 0.798±0.008 |
| 1.4 | 0.130±0.001 | 0.086±0.001 | 0.815±0.008 |
| 2 | 0.125±0.001 | 0.088±0.001 | 0.837±0.007 |
| 3 | 0.122±0.001 | 0.087±0.001 | 0.844±0.007 |
| 4 | 0.118±0.001 | 0.089±0.001 | 0.866±0.007 |
| **Analysis interval $\Delta t$ (minutes)** | *For sensitivity analysis (Fig. 5b), $\delta t$ held constant at 2 seconds* | | |
| 0.5 | 0.124±0.001 | 0.087±0.001 | 0.839±0.007 |
| 1 | 0.125±0.001 | 0.088±0.001 | 0.837±0.007 |
| 2 | 0.127±0.001 | 0.088±0.001 | 0.833±0.008 |
| 5 | 0.128±0.001 | 0.089±0.001 | 0.834±0.010 |
| 10 | 0.129±0.002 | 0.089±0.001 | 0.832±0.013 |
| 30 | 0.128±0.003 | 0.089±0.002 | 0.831±0.016 |
| **Diurnal time interval** | *For sensitivity analysis (Fig. 5c), $\delta t$ and $\Delta t$ held constant at 2 seconds and 1 minute, respectively* | | |
| 12-14h | 0.123±0.001 | 0.085±0.001 | 0.831±0.010 |
| 14-16h | 0.129±0.001 | 0.095±0.001 | 0.859±0.010 |
| 16-18h | 0.123±0.002 | 0.093±0.001 | 0.872±0.011 |
| **Date interval** | *For sensitivity analysis (Fig. 5c), $\delta t$ and $\Delta t$ held constant at 2 seconds and 1 minute, respectively* | | |
| [1]May 15-19 | 0.125±0.001 | 0.083±0.001 | 0.815±0.012 |
| [2]May 23-28 | 0.124±0.001 | 0.091±0.001 | 0.857±0.010 |
| [3]June 1-4 | 0.125±0.002 | 0.091±0.002 | 0.854±0.015 |

[1]Median, 10th percentile, and 90th percentile grain diameters of surface particles for this date interval are $d_{50} = 0.346±0.053$ mm, $d_{10} = 0.181±0.017$ mm, and $d_{90} = 0.580±0.061$, respectively. [2]For this date interval, reference grain diameters are $d_{50} = 0.417±0.056$, $d_{10} = 0.193±0.032$, and $d_{90} = 0.664±0.051$. [3]For this date interval, reference grain diameters are $d_{50} = 0.415±0.074$, $d_{10} = 0.194±0.029$, and $d_{90} = 0.677±0.079$.

Third, other factors, such as soil moisture [*Stout*, 2004; *Wiggs et al.*, 2004a; *Davidson-Arnott et al.*, 2005, 2008] and turbulence properties [*McKenna Neuman et al.*, 2000; *Davidson-Arnott et al.*, 2005] could have contributed to observed threshold variations. To investigate these factors, we perform a sensitivity analysis by time of day, which we interpret as a proxy for changes in soil moisture and atmospheric stability. This sensitivity analysis shows no systematic trend of increasing or decreasing thresholds through the course of the day (Fig. 5c). Though the three diurnal periods do show some variation in limiting fluid and impact threshold values, a similar ratio of fluid and impact thresholds is maintained throughout the day.

Fourth, unexplained variation in grain size [*Wiggs et al.*, 2004b; *Stout*, 2007] could have also played a role in the observed variations in effective threshold. To investigate the effect of grain size, we perform a sensitivity analysis by date, which we interpret as a proxy for the coarsening of surface grain size distributions through time at the Oceano field site (Table 2). This analysis indeed indicates slight increases in thresholds during periods of coarser sediment sizes, consistent with expectations (Fig. 5d). More generally, differences in grain size distributions do produce differences in effective thresholds among sites (Table 1), and these differences are consistent with expected impact and fluid thresholds [e.g., *Kok et al.*, 2012] for the soil particle size distributions at each site. Nonetheless, further work is needed to understand the sensitivity of thresholds to atmospheric stability [e.g., *Frank and Kocurek*, 1994], soil texture [e.g., *Greeley and Iversen*, 1985], and soil moisture [e.g., *Davidson-Arnott et al.*, 2008].

Fifth, it is also possible that increasing momentum extraction by the saltation cloud with increasing saltation intensity [e.g., *Sherman*, 1992] could bias our calculations, which assume constant roughness height $z_0$ (Eq. 3). Though we find no systematic variation in effective roughness height with transport activity $f_Q$ (Fig. 3b), it is nonetheless possible that, by using $z_0$ instead of the larger $z_s$, we systematically underestimate $u_{*,th}$ in comparison to actual values. It is also possible that the effective von Karman parameter increases during saltation [*Li et al.*, 2010], causing a contrasting systematic overestimation in $u_{*,th}$. However, based on a lack of variation in $z_s$ with $f_Q$, we expect no significant effect of wind momentum extraction by saltators, which would be required to cause a systematic change in the von Karman parameter $\kappa$ with $f_Q$.

### 5.2. Interpretations of dual thresholds
Recent wind tunnel measurements support our interpretations of how fluid and impact thresholds govern saltation transport. *Walter et al.* [2014] found that, so long as saltation activity is continuous (i.e., $f_Q = 1$), bed surface shear stress $\tau_0$ remains constant with changes in $Q$. By Owen's hypothesis [*Owen*, 1964], *Walter et al.* interpreted this constant $\tau_0$ as implying a constant rate of wind momentum dissipation at the bed surface equal to the impact threshold stress $\tau_{it}$. However, they also observed $\tau_0$ to decrease monotonically from $\tau_{ft}$ to $\tau_{it}$ during the transition from no saltation to continuous saltation [Fig. 4 in *Walter et al.*, 2014], suggesting that $\tau_0$, like effective threshold $\tau_{th}$ (Fig. 4a), is governed by changes in $f_Q$ modulating the relative importance of fluid and impact thresholds. *Paterna et al.* [2016] further observed that transport initiation events during weak saltation are dominantly related to energetic turbulent eddies directly mobilizing particles from the bed surface (i.e., fluid entrainment). In contrast, as saltation strengthens, variations in saltation flux decouple from the occurrence of turbulence structures, as the splash process (i.e., impact entrainment) plays an increasing role in sustaining saltation and transporting momentum to the bed surface. These findings, that the processes governing wind momentum transfer [*Paterna et al.*, 2016] and dissipation [*Walter et al.*, 2014] display fluid threshold control during weak saltation and impact threshold control during intense saltation, are consistent with our dual threshold hypothesis (Eq. 1) relating fluid to impact threshold contributions to the saltation activity.

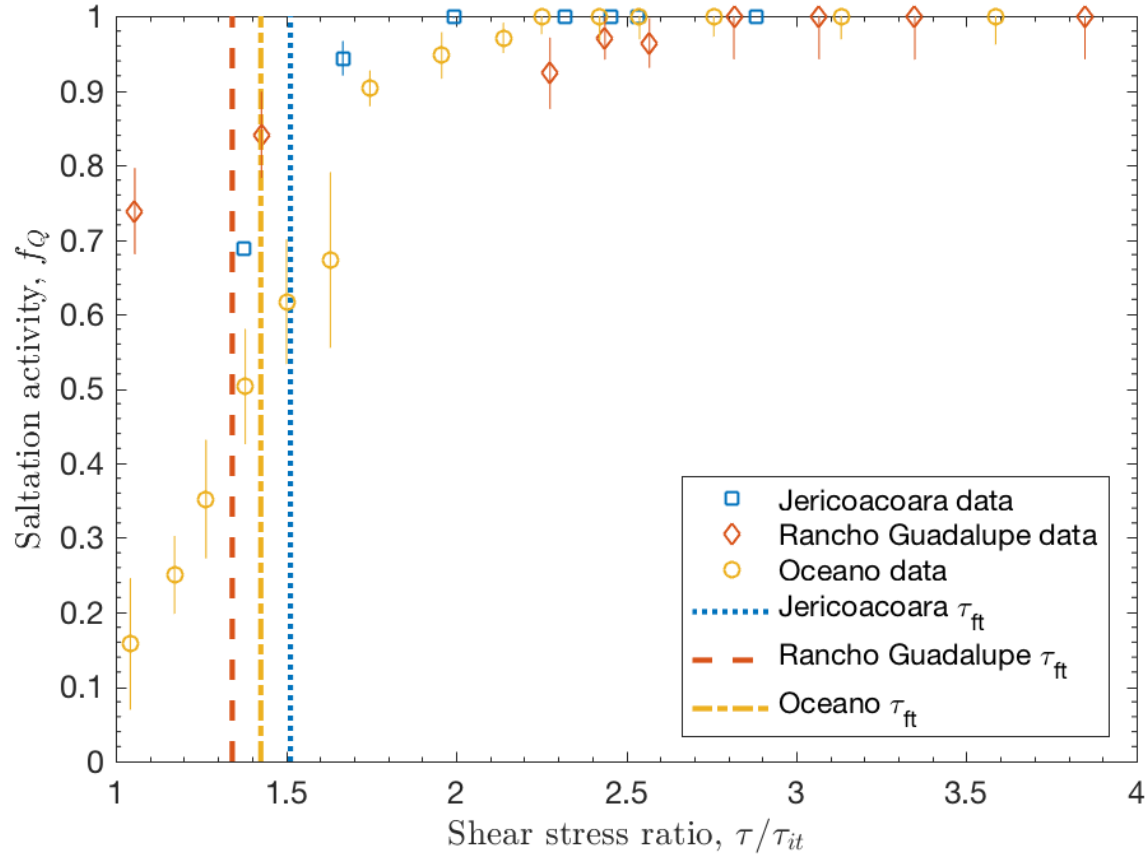

**Figure 6.** Saltation transport activity $f_Q$ versus shear stress ratio $\tau/\tau_{it}$. Shear stress values are computed over 30-minute intervals from the Reynolds stress [*Martin and Kok*, 2017], and corresponding $f_Q$ values are combined into bins by range of $\tau$. Error bars indicate standard error for each bin. Dashed lines indicate fluid threshold $\tau_{ft}$ at each site from Fig. 4.

Based on Eq. 1 and these recent wind tunnel studies, we expect impact threshold to dominate saltation dynamics in the limit of continuous ($f_Q = 1$) transport associated with moderate to strong saltation. We can gain insight into the importance of this continuous transport impact threshold control in saltation flux modeling by comparing fluid and impact threshold stresses calculated here to independent "flux-based" threshold stresses $\tau_{th,flux}$. These were obtained by *Martin and Kok* [2017] from the zero-intercept of the linear fit to saltation flux $Q$ versus shear stress $\tau$, i.e.,

$$Q \propto (\tau - \tau_{th,flux}). \tag{10}$$

For all three sites, values for $\tau_{th,flux}$ agree with our $\tau_{it}$ measurements within 2σ uncertainty ranges, and $\tau_{th,flux}$ values are closer to $\tau_{it}$ than to $\tau_{ft}$ (Fig. 4). We explain this general agreement between $\tau_{th,flux}$ and $\tau_{it}$ by the fact that most observed saltation at our field sites occurred under continuous or near-continuous transport conditions, for which $\tau_{it}$ dominantly controls wind momentum dissipation [*Walter et al.*, 2014] and the zero-intercept of the flux law [*Martin and Kok*, 2017]. Compared to the range of $\tau/\tau_{it} = \sim 1 - 4$ over which saltation is observed at our field sites and those reported in the literature [*Greeley et al.*, 1996; *Namikas*, 2003; *Li et al.*, 2010; *Farrell et al.*, 2012], the range of shear stresses for which saltation is intermittent (roughly $\tau/\tau_{it} = 1 - 1.5$, Fig. 6) is relatively narrow. We thus expect most saltation occurrence (and even more of the saltation flux, which increases with $\tau/\tau_{it}$), to be associated with continuous conditions and thus governed by $\tau_{it}$. Therefore, while $\tau_{ft}$ may merely serve as a rough boundary between intermittent and continuous saltation (Fig. 6), $\tau_{it}$ alone can serve as the *de facto* saltation threshold in studies modeling saltation flux over long time intervals (i.e., 30 minutes).

### 5.3. Selection of thresholds for aeolian saltation flux and dust emission modeling

Our findings offer insight into how thresholds should be incorporated into models for saltation occurrence, saltation flux, and dust emission. Both fluid and impact threshold should be used together when considering high-frequency saltation fluctuations, for which the flux-based threshold appears to be partially governed by averaging timescale [*Martin et al.*, 2013]. Dual thresholds should also be considered for modeling saltation on Mars [*Kok*, 2010b] or other planetary bodies [*Pähtz and Durán*, 2017] where the gap between fluid and impact thresholds is much larger than on Earth. However, when modeling saltation flux over longer analysis intervals (i.e., 30 minutes), such as in large-scale models for wind erosion [e.g., *Shao and Leslie*, 1997], dune migration [e.g., *Fryberger et al.*, 1979], and dust emission [e.g., *Gillette and Passi*, 1988], our results indicate that use of the impact threshold alone is sufficient. For the $\tau/\tau_{it} \approx 1 - 4$ range over which we observe saltation (Fig. 6), the choice of impact versus fluid threshold for saltation modeling (i.e., by Eq. 10) will produce at least a 10% difference in the prediction for saltation flux (and a >50% difference for $\tau/\tau_{it} < 2$). Therefore, adoption of the impact threshold, instead of the commonly-used higher fluid threshold [e.g., *Iversen and White*, 1982; *Marticorena and Bergametti*, 1995; *Shao*, 2008], could improve predictions for saltation flux. Since the dust emission flux is generally modeled as proportional to the saltation flux [e.g., *Marticorena and Bergametti*, 1995; *Shao*, 2008; *Kok et al.*, 2012], our findings have the potential to also improve predictions of dust emission fluxes.

Ideally, $\tau_{ft}$ and $\tau_{it}$ in saltation and dust modeling studies could be determined empirically by the methods we have described above, thus accounting for idiosyncratic variations in soil properties among sites [e.g., *Webb et al.*, 2016]. Otherwise, in cases where direct measurements of $\tau_{ft}$ and $\tau_{it}$ are not possible, such as on other planetary bodies [e.g., *Pähtz and Durán*, 2017], thresholds could be estimated from theoretical relationships [e.g., *Claudin and Andreotti*, 2006; *Kok*, 2010b], numerical models [e.g., *Almeida et al.*, 2008; *Kok*, 2010a; *Pähtz et al.*, 2012], or experimental studies [e.g., *Chepil*, 1945; *Iversen and White*, 1982], based on known median surface grain diameters and atmospheric properties. However, such estimates of threshold values and their effects on saltation occurrence may be complicated by uncertainties in measurements of wind speeds, turbulence properties, soil moisture, and sediment size distributions. Further work is needed to improve these measurements and to constrain their relationships to threshold values.

## 6. Conclusions

Here we offered the first field-based evidence of distinct fluid and impact thresholds for the respective initiation and cessation of aeolian saltation. The measured ratio of these thresholds is consistent with past laboratory and numerical studies of dual thresholds. We calculated fluid and impact thresholds by examining their roles in regulating saltation activity. When saltation is active a small fraction of the time, wind exceedance of fluid threshold controls saltation occurrence. As saltation activity increases, so too does the influence of the impact threshold, until saltation occurrence is controlled mostly by wind exceedance of impact threshold under near-continuous transport conditions. Although dual thresholds are thus required for certain saltation modeling applications, our results indicate that the impact threshold alone is sufficient for predicting the time-averaged (~30 minute) saltation flux on Earth. Consequently, we suggest that parameterizations of sand and dust transport in aeolian process models, which currently predominantly use the fluid threshold, should instead adopt the impact threshold for predictions.

**Appendix A. Accounting for false negatives in calculation of saltation activity**

Here, we derive a method to account for the possibility of "false negatives" (i.e., instances in which saltation transport occurs but is not detected) when computing saltation activity $f_Q$. To do so, we distinguish the measured saltation detection rate $f_D$ from the actual saltation activity $f_Q$. We compute $f_D$ as the fraction of averaging intervals $\delta t$ in each analysis interval $\Delta t$ for which total particle counts rate $N$ is nonzero. We then estimate $f_Q$ from $f_D$ by calculating the expected rate of false negatives for particle arrivals occurring as a Poisson counting process. We detail this procedure below.

By Bayes Theorem, we have:
$$f_{D|Q} = \frac{f_D f_{Q|D}}{f_Q}, \tag{A1}$$
where $f_{D|Q}$ is the conditional probability of detecting transport when it is active, and $f_{Q|D}$ is the probability that transport is actually active when it has been detected. We observed that the Wenglor optical particle counters did not produce "false positives," i.e. that detection necessarily implied transport and therefore that $f_{Q|D} = 1$. However, we found that during conditions of weak transport or few Wenglor counters, false negatives could occur with some regularity due to the limited sampling volume of individual counters. Denoting these false negatives as $f_{\sim D|Q}$ and noting that $f_{\sim D|Q} + f_{D|Q} = 1$, we can restate Eq. A1 as:
$$f_Q = \frac{f_D}{1 - f_{\sim D|Q}}. \tag{A2}$$
To estimate the rate of false negatives $f_{\sim D|Q}$, we treat particle arrivals as a Poisson counting process. For such a process:
$$f_{\sim D|Q} = \exp(-\lambda), \tag{A3}$$
where $\lambda$ is the average arrival rate of particles per $\delta t$ averaging interval at times when transport is active. We calculate $\lambda$ as:
$$\lambda = \bar{N} \delta t / f_D, \tag{A4}$$
where $\bar{N}$ is the mean particle counts rate during the analysis interval $\Delta t = 1$ minute. Combining Eqs. A2-A4, we have:
$$f_Q = \frac{f_D}{1 - \exp(-\lambda)} = \frac{f_D}{1 - \exp(-\bar{N} \delta t / f_D)}. \tag{A5}$$

The effect of the false negative correction in computing $f_Q$ can be seen in Supporting Information Fig. S1. During strong transport, $\lambda$ is large so $f_Q \approx f_D$ in Eq. A5. However, when transport is weak, $\delta t$ is small, or the number of Wenglors are few, $\lambda$ can be much smaller than 1, and therefore the correction causes $f_Q$ to significantly exceed $f_D$.

# Appendix B. Estimation of uncertainties for effective threshold wind speeds, shear velocities, and shear stresses

To facilitate uncertainty estimation for computed effective thresholds, we combine effective threshold wind speed values $u_{th}$ from individual $\Delta t$ analysis intervals into bins defined by ranges of flux activity $f_Q$. For each bin, we compute bin-averaged threshold and activity values for $u_{th}$ and $f_Q$, then we calculate their uncertainties from the standard errors of values in each bin. When converting each bin-averaged effective threshold wind speed $u_{th}$ to an effective threshold shear velocity $u_{*,th}$ and shear stress $\tau_{th}$, we perform error propagation to estimate uncertainties in $u_{*,th}$ and $\tau_{th}$.

For each site, we group values of $u_{th}$ for individual $\Delta t$ together into bins covering ranges of $f_Q$. To accommodate the uneven spread of data points across the range of possible $f_Q$, we allow for creation of bins covering varying ranges of $f_Q$. We establish these criteria to balance the need for a sufficient number of data points in each bin with the need to limit the maximum width of the bins. The procedure for generating the binned values for each site is as follows:
(1) Sort all $u_{th}$ data points in order of increasing $f_Q$. Because the effective threshold calculation (Eq. 2) assumes intermittent transport conditions, exclude data points with $f_Q < 0.05$ and $f_Q > 0.95$.
(2) Starting from the lowest remaining $f_Q$, add data points to the bin, until the following criteria are achieved for the bin
   a. $\max(f_Q) - \min(f_Q) \geq 0.1$ (minimum bin width), AND
   b. There are at least 3 points in the bin OR $\max(f_Q) - \min(f_Q) > 0.2$ (maximum bin width).
(3) Once the bin is full, repeat step 2 for the next bin.

For each bin $i$, we determine the mean value for flux activity $f_{Q,i}$ and its uncertainty $\sigma_{f_{Q,i}}$ as:

$$f_{Q,i} = \sum_j f_{Q,j}/N_i, \tag{B1}$$

$$\sigma_{f_{Q,i}} = \frac{\sqrt{\sum_j (f_{Q,j} - f_{Q,i})^2}}{\sqrt{N_i}}, \tag{B2}$$

where $f_{Q,j}$ are the individual values of flux activity in the bin and $N_i$ is the total number of values in the bin. Eq. B2 is computed based on the typical formulation for the standard error [Eq. 4.14 in *Bevington and Robinson*, 2003]. Similarly, the mean effective threshold wind speed $u_{th,i}$ and its uncertainty $\sigma_{u_{th,i}}$ are:

$$u_{th,i} = \sum_j u_{th,j}/N_i, \tag{B3}$$

$$\sigma_{u_{th,i}} = \frac{\sqrt{\sum_j (u_{th,j} - u_{th,i})^2}}{\sqrt{N_i}}, \tag{B4}$$

where $u_{th,j}$ are the individual values of effective threshold in the bin.

We convert binned values for effective threshold wind speed $u_{th,i}$ to effective threshold shear velocity $u_{*,th}$ by Eq. 3 and to effective threshold stress $\tau_{th}$ by Eq. 4. Using the error propagation formula [Eq. 3.14 in *Bevington and Robinson*, 2003], we propagate effective threshold wind speed uncertainty $\sigma_{u_{th}}$ and roughness height uncertainty $\sigma_{\ln(z_0)}$ to derive uncertainty in threshold shear velocity $\sigma_{u_{*,th}}$ and threshold stress $\sigma_{\tau_{th}}$:

$$\sigma_{u_{*,th}} = \frac{u_{*,th}}{u_{th}} \sqrt{\sigma_{u_{th}}^2 + \frac{\ln^2(\sigma_{z_0})}{\ln^2(z_U/z_0)} u_{*,th}^2}, \tag{B5}$$

$$\sigma_{\tau_{th}} = 2\rho_f u_{*,th} \sigma_{u_{*,th}}. \tag{B6}$$

To determine uncertainties in fluid $\tau_{ft}$ and impact $\tau_{it}$ threshold stresses (Eqs. 7-8), we propagate uncertainties in the fitting intercept $a$ and fitting slope $b$:

$$\sigma_{\tau_{ft}} = \sigma_a, \tag{B7}$$

$$\sigma_{\tau_{it}} = \sqrt{\sigma_a^2 + \sigma_b^2 + 2\sigma_{ab}^2}. \tag{B8}$$

where $\sigma_a$ and $\sigma_b$ are the respective uncertainties in $a$ and $b$, and $\sigma_{ab}^2$ is their covariance. We then apply error propagation to the threshold ratio $u_{*it}/u_{*ft}$ to calculate its uncertainty:

$$\sigma_{u_{*it}/u_{*ft}} = \frac{1}{2} \sqrt{\frac{\sigma_{\tau_{it}}^2}{\tau_{it}\tau_{ft}} + \frac{\sigma_{\tau_{ft}}^2 \tau_{it}}{\sigma_{\tau_{ft}}^3}}. \tag{B9}$$

## Appendix C: List of variables
Below, we list all variables described in the manuscript. Typical units for variables are given in parentheses, if applicable.

$u_*$ = shear velocity (m/s)
$u_{*ft}$ = fluid threshold shear velocity (m/s)
$u_{*it}$ = impact threshold shear velocity (m/s)
$u_{*it}/u_{*ft}$ = shear velocity threshold ratio
$u_{*,th}$ = effective threshold shear velocity (m/s)
$u_{*,Re}$ = Reynolds stress-based shear velocity (m/s)
$\tau$ = wind shear stress (Pa)
$\tau_{ft}$ = fluid threshold shear stress (Pa)
$\tau_{it}$ = impact threshold shear stress (Pa)
$\tau_{th}$ = effective threshold shear stress (Pa)
$f_Q$ = saltation transport "activity"; i.e., fraction of time saltation is active
$u(t)$ = time series of horizontal wind speed (m/s)
$z_U$ = anemometer height above the sand surface (m)
$u_{ft}$ = fluid threshold wind speed (m/s)
$u_{it}$ = impact threshold wind speed (m/s)
$u_{th}$ = effective threshold wind speed (m/s)
$d_{50}$, median diameter of surface particles by volume (mm)
$d_{10}$, 10$^{th}$ percentile diameter of surface particles by volume (mm)
$d_{90}$, 90$^{th}$ percentile diameter of surface particles by volume (mm)
$N$ = vertically integrated saltation particle counts rate (counts/s)
$\delta t$ = averaging time interval (s)
$\Delta t$ = analysis time interval (minutes)
$\Phi_u$ = cumulative distribution of streamwise wind speed $u$
$\kappa$ = von Karman parameter
$z_0$ = aerodynamic roughness height (m)
$\sigma_{\ln(z_0)}$ = natural log uncertainty in aerodynamic roughness height
$z_s$ = effective roughness height, accounting for saltation-induced roughness (m)
$\rho_f$ = air density (kg/m$^3$)
$\theta$ = angle of horizontal wind relative to dominant sand-transporting wind
$|z/L|$ = stability parameter
$Q$ = saltation flux (g/m/s)
$a$ = least squares linear fitting intercept
$b$ = least squares linear fitting slope
$\tau_0$ = bed surface shear stress (Pa)
$\tau_{th,flux}$ = flux-based estimate of threshold stress (Pa)
$f_D$ = saltation detection rate
$f_{Q|D}$ = probability that transport is actually active when it has been detected
$f_{D|Q}$ = conditional probability of detecting transport when it is active
$f_{\sim D|Q}$ = conditional probability of not detecting transport when it is active
$\lambda$ = average arrival rate of particles (counts/s)

$\overline{N}$ = mean vertically integrated saltation particle counts rate (counts/s)
$f_{Q,i}$ = mean value for saltation transport activity for bin $i$
$\sigma_{f_{Q,i}}$ = uncertainty in saltation transport activity for bin $i$
$f_{Q,j}$ = individual values $j$ of saltation activity in bin $i$
$N_i$ = number of values in bin $i$
$u_{th,i}$ = mean value for effective threshold wind speed for bin $i$ (m/s)
$\sigma_{u_{th,i}}$ = uncertainty in effective threshold wind speed for bin $i$ (m/s)
$u_{th,j}$ = individual values $j$ of effective threshold wind speed in bin $i$ (m/s)
$\sigma_{u_{*,th}}$ = uncertainty in effective threshold shear velocity (m/s)
$\sigma_{\tau_{th}}$ = uncertainty in effective threshold shear stress (Pa)
$\sigma_a$ = uncertainty of linear fitting intercept
$\sigma_b$ = uncertainty of linear fitting slope
$\sigma_{ab}^2$ = covariance of fitting slope and intercept
$\sigma_{\tau_{ft}}$ = uncertainty in fluid threshold stress (Pa)
$\sigma_{\tau_{it}}$ = uncertainty in impact threshold stress (Pa)
$\sigma_{u_{*it}/u_{*ft}}$ = uncertainty in shear velocity threshold ratio

**ACKNOWLEDGEMENTS.** U.S. National Science Foundation (NSF) Postdoctoral Fellowship EAR-1249918 to R.L.M. and NSF grant AGS-1358621 to J.F.K. supported this research. Research was also sponsored by the Army Research Laboratory and was accomplished under Grant Number W911NF-15-1-0417. The views and conclusions contained in this document are those of the authors and should not be interpreted as representing the official policies, either expressed or implied, of the Army Research Laboratory or the U.S. Government. The U.S. Government is authorized to reproduce and distribute reprints for Government purposes notwithstanding any copyright notation herein. Oceano Dunes State Vehicular Recreation Area, Rancho Guadalupe Dunes Preserve, and Jericoacoara National Park provided essential site access and support. Jericoacoara fieldwork is registered with the Brazilian Ministry of the Environment (#46254-1 to J. Ellis). We thank Marcelo Chamecki for advice on treatment of wind data, Chris Hugenholtz and Tom Barchyn for equipment help, Doug Jerolmack for lab access for grain-size analysis, and Jean Ellis, Paulo Sousa, Peter Li, Francis Turney, Arkayan Samaddar, and Livia Freire for field assistance. We also thank M. Bayani Cardenas and four anonymous reviewers for their thoughtful suggestions on several versions of this manuscript. Data included in the analysis for this paper can be found on the Zenodo data repository at http://doi.org/10.5281/zenodo.574896.

# References

Almeida, M. P., E. J. R. Parteli, J. S. Andrade, and H. J. Herrmann (2008), Giant Saltation on Mars, *Proc. Natl. Acad. Sci. U. S. A.*, *105*(17), 6222–6226, doi:10.1073/pnas.0800202105.

Anderson, R. S., and P. K. Haff (1988), Simulation of eolian saltation, *Science*, *241*(4867), 820–823, doi:10.1126/science.241.4867.820.

Ayoub, F., J.-P. Avouac, C. E. Newman, M. I. Richardson, A. Lucas, S. Leprince, and N. T. Bridges (2014), Threshold for sand mobility on Mars calibrated from seasonal variations of sand flux, *Nat. Commun.*, *5*, 5096.

Baas, A. C. W. (2008), Challenges in aeolian geomorphology: Investigating aeolian streamers, *Geomorphology*, *93*(1–2), 3–16, doi:10.1016/j.geomorph.2006.12.015.

Bagnold, R. A. (1937), The transport of sand by wind, *Geogr. J.*, *89*(5), 409–438, doi:10.2307/1786411.

Bagnold, R. A. (1941), *The Physics of Blown Sand and Desert Dunes*, Dover, London.

Barchyn, T. E., and C. H. Hugenholtz (2011), Comparison of four methods to calculate aeolian sediment transport threshold from field data: Implications for transport prediction and discussion of method evolution, *Geomorphology*, *129*(3–4), 190–203, doi:10.1016/j.geomorph.2011.01.022.

Barchyn, T. E., C. H. Hugenholtz, B. Li, C. McKenna Neuman, and R. S. Sanderson (2014a), From particle counts to flux: Wind tunnel testing and calibration of the "Wenglor" aeolian sediment transport sensor, *Aeolian Res.*, *15*, 311–318, doi:10.1016/j.aeolia.2014.06.009.

Barchyn, T. E., R. L. Martin, J. F. Kok, and C. H. Hugenholtz (2014b), Fundamental mismatches between measurements and models in aeolian sediment transport prediction: The role of small-scale variability, *Aeolian Res.*, *15*, 245–251, doi:10.1016/j.aeolia.2014.07.002.

Bevington, P. R., and D. K. Robinson (2003), *Data reduction and error analysis for the physical sciences*, 3rd ed., McGraw-Hill, New York.

van Boxel, J. H., G. Sterk, and S. M. Arens (2004), Sonic anemometers in aeolian sediment transport research, *Geomorphology*, *59*(1–4), 131–147, doi:10.1016/j.geomorph.2003.09.011.

Bridges, N. T., F. Ayoub, J.-P. Avouac, S. Leprince, A. Lucas, and S. Mattson (2012), Earth-like sand fluxes on Mars, *Nature*, *485*(7398), 339–342, doi:10.1038/nature11022.

Carneiro, M. V., K. R. Rasmussen, and H. J. Herrmann (2015), Bursts in discontinuous Aeolian saltation, *Sci. Rep.*, *5*, doi:10.1038/srep11109.
25

Chepil, W. S. (1945), Dynamics of Wind Erosion. 2. Initiation of soil movement, *Soil Sci.*, *60*(5), 397, doi:10.1097/00010694-194511000-00005.

Claudin, P., and B. Andreotti (2006), A scaling law for aeolian dunes on Mars, Venus, Earth, and for subaqueous ripples, *Earth Planet. Sci. Lett.*, *252*(1–2), 30–44, doi:10.1016/j.epsl.2006.09.004.

Creyssels, M., P. Dupont, A. O. El Moctar, A. Valance, I. Cantat, J. T. Jenkins, J. M. Pasini, and K. R. Rasmussen (2009), Saltating particles in a turbulent boundary layer: Experiment and theory, *J. Fluid Mech.*, *625*, 47–74, doi:10.1017/S0022112008005491.

Davidson-Arnott, R. G. D., K. MacQuarrie, and T. Aagaard (2005), The effect of wind gusts, moisture content and fetch length on sand transport on a beach, *Geomorphology*, *68*(1–2), 115–129, doi:10.1016/j.geomorph.2004.04.008.

Davidson-Arnott, R. G. D., Y. Yang, J. Ollerhead, P. A. Hesp, and I. J. Walker (2008), The effects of surface moisture on aeolian sediment transport threshold and mass flux on a beach, *Earth Surf. Process. Landf.*, *33*(1), 55–74, doi:10.1002/esp.1527.

Farrell, E. J., D. J. Sherman, J. T. Ellis, and B. Li (2012), Vertical distribution of grain size for wind blown sand, *Aeolian Res.*, *7*, 51–61, doi:10.1016/j.aeolia.2012.03.003.

Fletcher, B. (1976), The incipient motion of granular materials, *J. Phys. Appl. Phys.*, *9*(17), 2471, doi:10.1088/0022-3727/9/17/007.

Frank, A., and G. Kocurek (1994), Effects of atmospheric conditions on wind profiles and aeolian sand transport with an example from White Sands National Monument, *Earth Surf. Process. Landf.*, *19*(8), 735–745, doi:10.1002/esp.3290190806.

Fryberger, S. G., G. Dean, and E. D. McKee (1979), Dune forms and wind regime, in *A Study of Global Sand Seas, U.S. Geological Survey Professional Paper 1052, E. D. McKee, Ed.*, pp. 137–170.

Gillette, D. A., and R. Passi (1988), Modeling dust emission caused by wind erosion, *J. Geophys. Res. Atmospheres*, *93*(D11), 14233–14242, doi:10.1029/JD093iD11p14233.

Greeley, R., and J. D. Iversen (1985), *Wind as a Geological Process on Earth, Mars, Venus and Titan*, Cambridge University Press, New York.

Greeley, R., D. G. Blumberg, and S. H. Williams (1996), Field measurements of the flux and speed of wind-blown sand, *Sedimentology*, *43*(1), 41–52, doi:10.1111/j.1365-3091.1996.tb01458.x.

Guala, M., M. Metzger, and B. J. McKeon (2011), Interactions within the turbulent boundary layer at high Reynolds number, *J. Fluid Mech.*, *666*, 573–604, doi:10.1017/S0022112010004544.

Supporting Information for

# Field measurements demonstrate distinct initiation and cessation thresholds governing aeolian sediment transport flux

Raleigh L. Martin[1] and Jasper F. Kok[1]

[1]Department of Atmospheric and Oceanic Sciences, University of California, Los Angeles, CA 90095

**Contents of this file**

    Figure S1

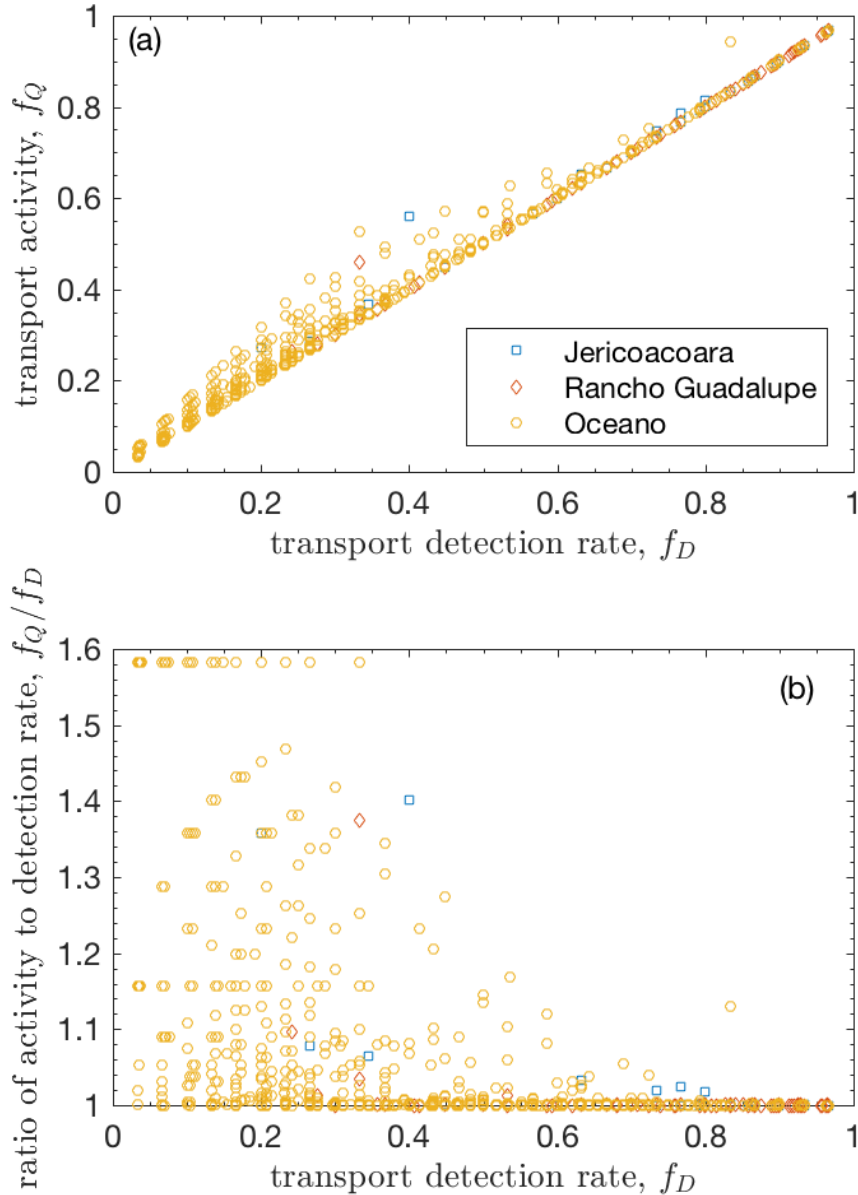

**Figure S1.** (a) Comparison of saltation transport activity $f_Q$, which includes correction for estimated rate of false negatives (Eq. A5), to the actual transport detection rate $f_D$. (b) Ratio of transport activity to detection rate $f_Q/f_D$ versus detection rate $f_D$, illustrating the relative magnitude of the correction for false negatives. This correction tends to be strongest when transport is weak, due to the higher probability of false negatives. The difference between $f_Q$ and $f_D$ is also affected by changes in the background particle detection rate, which can be affected by the number and height of Wenglor detectors and the nature of transport fluctuations.